\documentclass[
prl
,showpacs
,showkeys
,aps
,amsfonts
,amssymb
,byrevtex
,pdftex  
,twocolumn,nofootinbib
]{revtex4-1}

\usepackage{graphicx}  
\usepackage[titletoc]{appendix}
\usepackage{amsmath, amssymb, bm}   
\usepackage[dvips]{color}
\usepackage{natbib}
\usepackage{hyperref}

\def\){\right)} 
\def\({\left(} 
\def\]{\right]} 
\def\[{\left[}

\begin{document}

\title{
Non-radial Oscillation Modes of Compact Stars with a Crust}

\author{%
Cesar V\'asquez Flores
}
\email{cesarovfsky@gmail.com}

\affiliation{Departamento de F\'isica, Universidade Federal do  Maranh\~ao,Campus Universit\'ario do Bacanga, CEP 65080-805, S\~ao Lu\'is, Maranh\~ao, Brazil}

\author{%
Zack B. Hall II
}

\affiliation{Department of Physics and Astronomy, University of North Carolina, Chapel Hill, NC 27599, USA}

\author{%
Prashanth Jaikumar
}

\email{prashanth.jaikumar@csulb.edu}

\affiliation{Department of Physics $\&$ Astronomy,
California State University Long Beach, Long Beach, CA
  90840, U.S.A. }


\begin{abstract}

Oscillation modes of isolated compact stars can, in principle, be a fingerprint of the equation of state (EoS) of dense matter. We study the non-radial high-frequency $l$=2 spheroidal modes of neutron stars and strange quark stars, adopting a two-component model (core and crust) for these two types of stars. Using perturbed fluid equations in the relativistic Cowling approximation, we explore the effect of a strangelet or hadronic crust on the oscillation modes of strange stars. The results differ from the case of neutron stars with a crust. In comparison to fluid-only configurations, we find that a solid crust on top of a neutron star increases the $p$-mode frequency slightly with little effect on the $f$-mode frequency, whereas for strange stars, a strangelet crust on top of a quark core significantly increases the $f$-mode frequency with little effect on the $p$-mode frequency.  

\end{abstract}



\maketitle

\section{I. Introduction}
Advanced LIGO has ushered in a new era of compact star observations with the recent direct detections of gravitational waves from the inspiral and merger of binary Black Holes~\cite{Abbott:2016blz,Abbott:2016nmj,Abbott:2017vtc}. While the study of electromagnetic radiation from the surface of neutron stars already yields information on the state of high density matter inside~\cite{Steiner:2010fz,Read:2008iy,Ozel:2010fw,Page:2010aw,Newton:2013zaa,Negreiros:2013ina,Miller:2011wt,Guillot:2014lla}, gravitational waveforms arising from quadrupolar deformations of the neutron star due to tidal effects, vibrations, rotation or elastic strain in the crust can provide additional constraints~\cite{Agathos:2015uaa,Jones:2001ui}. With current sensitivities, detectable signals come from transient but violent events such as the merger and ringdown of colliding neutron star and black hole binaries~\cite{Creighton:2003nm,Bauswein:2012ya}. Weaker signals are expected from oscillations of isolated compact stars~\cite{Kokkotas:1999bd,Kokkotas:1999mn}, sustained through instabilities~\cite{Stergioulas:2003yp} or at least excited temporarily before they are damped by fluid viscosity and gravitational waves. The classification of these modes and their study is a mature field~\cite{Chirenti:2012wn,Miniutti:2002bh,Detweiler:1985zz,PenaGaray:2008qe,Sotani:2013jya}.

Focusing on non-radial modes of oscillation in isolated compact stars, the even-parity or spheroidal modes arise from density and pressure perturbations to the star, while odd-parity axial modes are non-trivial only for rotating stars. Lugones and V\'asquez Flores~\cite{Flores:2013yqa} recently compared the spectrum of spheroidal $f,p,g$ modes for hadronic, strange and hybrid stars, in order to find discriminating features among them. All these modes couple to gravitational radiation, hence it is important to determine the corresponding (complex) oscillation frequencies. From their conclusion for the $f$-mode, it appears difficult (though possible in some cases) to discern hadronic stars from strange stars around the observed mass range of 1.4-2$M_{\odot}$. However, for the first $p$-mode, there is a wide separation in the frequencies for the hadronic and strange stars in this mass range. One can potentially exploit this feature of the spectrum to constrain the mass and radius of the compact star independently. The $g$-mode arises when considering effects of chemical inhomogeneity, non-zero temperature or discontinuity between two otherwise homogeneous phases~\cite{Miniutti:2002bh}. Lugones and V\'asquez Flores~\cite{Flores:2013yqa} conclude that the quark-hadron discontinuity in hybrid stars leads to higher frequency $g$-modes than those driven by chemical composition or temperature effects. Mapping out the oscillation spectrum is clearly important as a first step towards more complex calculations in numerical relativity that can serve as templates for gravitational waveforms in detectors~\cite{Kokkotas:1999bd}. Several other works have also discussed the differences in mode frequencies between neutron stars and strange/hybrid stars~\cite{Kojima:2002iv,Moraes:2014dra,Chatziioannou:2015uea}, adopting a stellar model with homogeneous phases. 

The main aim of this paper is to consider the effect of a crust on the spectrum of spheroidal modes in a neutron or strange star, and compare to modes for homogeneous, zero-temperature stars. While the crust is only a small fraction of the total stellar mass, it can change the oscillation frequencies of the spheroidal and toroidal modes from the homogeneous case. Toroidal and core-crust interface modes have also been connected with quasi-periodic oscillations in magnetar flares~\cite{Watts:2006mr} and Gamma-Ray Burst (GRB) precursors~\cite{Tsang:2011ad} respectively. Our 2-component model for neutron stars includes a BPS crust, while for strange stars, we consider a crust made of strangelets~\cite{Jaikumar:2005ne,Alford:2006bx} or a thin hadronic crust. This extends the scope of previous works where homogeneous phases were considered~\cite{Flores:2013yqa,Sotani:2013jya,Kojima:2002iv}.  Our interest is in the effects of a crust for two reasons: one is that whereas isolated neutron stars are expected to have a solid crust, only a few works address its effects on the $f,p$ modes for neutron stars in the {\it relativistic} approximation, e.g.,~\cite{Yoshida:2002vd}. 
Original works such as~\cite{McDermott:1988,Strohmayer} found very little modification of the $f,p$ modes due to the crust, but it is worth revisiting this problem since these works typically use an old parameterization of the equation of state, and employ a Newtonian approximation. Our present work is the first to study the effect of a crust on the non-radial modes in strange stars, including relativistic effects. Another reason for studying crustal effects is the finding from an exploration of the toroidal $r$-modes~\cite{Rupak:2013dxa,Rupak:2010qg}: if strange stars are to be a viable model for rapidly rotating compact stars, given the observed spin frequencies~\cite{Chakrabarty:2008gz} and a maximum mass around 2$M_{\odot}$, a crystalline (superconducting) quark crust on top of the homogenous superconducting quark phase is needed.  We would naturally like to ask if and how spheroidal modes of bare strange stars are modified by a crust, although in this paper, we limit ourselves to non-superconducting quark matter, leaving the crystalline supersolid crust to a future study. In this context, Mannarelli et al.~\cite{Mannarelli:2015jia} have recently explored torsional oscillations of strange stars with such a supersolid crust.
Since we discuss strange stars, it is pertinent that Quantum Chromodynamics (QCD) at large baryon density and low temperature favors a maximally symmetric phase of homogeneous superconducting quark matter called the color-flavor-locked (CFL) phase~\cite{Alford:1998mk}. Stellar oscillations in CFL matter have only just started to receive attention~\cite{Flores:2017hpb}. However, at compact star densities, the phase of matter is less certain. For example, the core may be in the kaon-condensed CFL phase while the crust can be in the crystalline superconducting phase~\cite{Mannarelli:2006fy}. By adopting a 2-component model that can be extended to color superconductivity in the core and crust of the strange star, we hope to get closer to realistic models of such stars.

We proceed in sec II to describe the EoS used in this paper to model self-bound strange stars with a quark crust as well as neutron stars with a BPS crust. In sec III, we use these models to calculate the $f,p$-mode frequencies and compare to results for homogeneous strange stars and neutron stars. In sec IV, we summarize our conclusions in the context of discerning strange quark stars from neutron stars as gravitational wave sources. The appendix contains the oscillation equations in the relativistic Cowling approximation that we solve to find the oscillation modes.
\section{II. Equation of state : core and crust}
In this section, we present the EoS used in modeling the compact star (either neutron or strange star) with a crust. The mass $M$ and radius $R$ are  determined by solving the Tolman-Oppenheimer-Volkov (TOV) equations with this EoS, which also provides the background for the perturbations. 
\subsection{Strange Quark Stars} 
\begin{figure}[htbp]
\begin{center}
\includegraphics[height=3.2in,width=2.5in,angle=270]{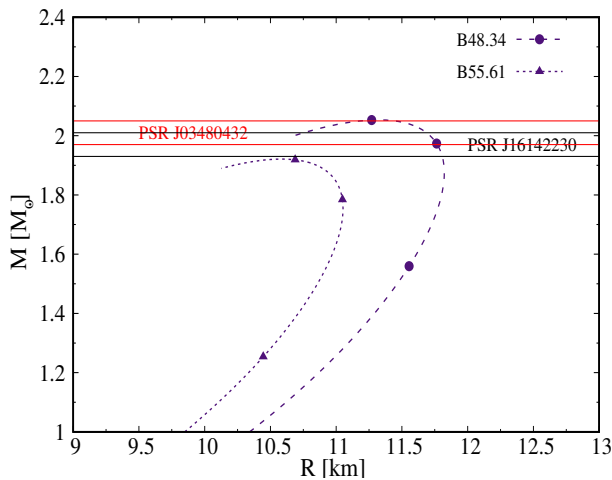}
\caption{Mass-Radius relation for homogeneous quark stars described by the EoS (Eq.\ref{EoS-1}) with $B$=48.34 MeV/fm$^3$ ($\mu_{\rm crit}$=300 MeV) and $B$=55.61 MeV/fm$^3$ ($\mu_{\rm crit}$=310 MeV) . The interaction parameter $a_4$=0.7 and $m_s$=100 MeV. Measured limits on the mass of PSR J1614-2230 (1.97$\pm$0.04 $M_{\odot}$) and PSR J0348-0432 (2.01$\pm$0.04 $M_{\odot}$) are shown.}
\label{mrbare}
\end{center}
\end{figure}

{\it Core EoS:} The core is assumed to be comprised of homogeneous, charge neutral 3-flavor interacting quark matter. For simplicity, we describe this phase using the simple thermodynamic Bag model EoS~\cite{Alford:2004pf} with ${\cal O}(m_s^4)$ corrections that account for the moderately heavy strange quark. Perturbative interactions to the pressure $P_q$ of non-interacting quark matter~\cite{Fraga:2001id} may be subsumed into a parameter $(1-a_4)\sim {\cal O}(\alpha_s^2)\approx 0.3$ as suggested in~\cite{Alford:2004pf}, extending the applicability of the model to stars as heavy as $\approx 2M_{\odot}$. The core EoS is~\cite{Asbell:2017zxp}
\begin{align}
\label{EoS-1}
P_{q,\mathrm{core}}&=\frac{1}{3}(\epsilon-4 B)-\frac{m_s^2}{3\pi}\sqrt{\frac{\epsilon-B}{a_4}}\nonumber\\
&+\frac{m_s^4}{12\pi^2}\[2 - \frac{1}{a_4}+3\ln\left(\frac{8\pi}{3m_s^2}\sqrt{\frac{\epsilon-B}{a_4}}\right)\] , 
\end{align}
where $\epsilon$ is the energy density of homogeneous quark matter (also to ${\cal O}(m_s^4)$ in the Bag model). $B$ is the Bag constant, fixed by requiring that the first-order transition between neutral quark matter and the vacuum ($P$=0) occur at a quark chemical potential $\mu_q$=$\mu_{\rm crit} < m_N/3$~\cite{Alford:2006bx}. This is consistent with the hypothesis of absolute stability for 3-flavor quark matter, meaning such matter is self-bound. Corrections due to the superconducting gap $\Delta$ may be included in the EoS as in~\cite{Rupak:2013dxa}, but we prefer to avoid a proliferation of parameters in this initial study. In any case, effects of the quark BCS gap on the oscillation spectrum of homogeneous strange stars have been studied recently~\cite{Flores:2017hpb}. Fig.\ref{mrbare} shows the mass-radius relationship for homogeneous quark stars for different values of the Bag constant $B$. 
\\
{\it Crust EoS:} A two-component structure for strange quark stars was suggested in~\cite{Jaikumar:2005ne} and developed further in subsequent works~\cite{Alford:2006bx,Alford:2008ge}. Models of strange stars that consist entirely of homogeneous strange quark matter or a thin nuclear crust suspended on top have large quark density as the quark surface is approached and the pressure goes to zero. They also predict large electric fields at the surface (or just below, in the case of a nuclear crust). The two-component model we use differs from these in that it considers a heterogeneous crust on top of fluid quark matter. Relaxing the condition of local charge neutrality allows quarks and electrons to form a mixed phase, with the result that it smooths the density gradient and produces negligible electric fields at the star's surface. Short-range (screened) electric fields inside the mixed phase can be tolerated if the Gibbs free energy is lowered sufficiently. The thickness of such a crust would be small but on the same order as the crust of a neutron star, approximately $10^4$ cm, and amount to no more than 1\% of the stellar radius. Assuming that the crust is in this non-superconducting, globally neutral mixed phase of (positively charged) strangelets and electrons~\footnote{We include QCD-inspired corrections for the strangelets as with the homogeneous phase~\cite{Fraga:2001id}.}, the composition of the crust changes with depth as the quark phase fraction $x$ increases from zero at the surface to one in the homogeneous phase\footnote{This chemical composition gradient can lead to crustal $g$-modes which we do not calculate in this work, but have been discussed for neutron stars in~\cite{1987MNRAS.227..265F}.}. Phase coexistence (neglecting momentarily the contribution from surface tension) with stable strangelets requires that  the pressure inside and outside the strangelet be the same, which means that the quark pressure $P_q(\mu_q,\mu_e)$ in the mixed phase is zero, so that the pressure in the crust is only due to electrons. It is given by~\cite{Jaikumar:2005ne}
\begin{eqnarray}
\label{pmix}
&&P_{\rm crust}=\frac{\tilde{\mu}_e^4}{12\pi^2}\,,\\
&&\tilde{\mu}_e=\frac{n_Q}{\chi_Q}\left(1-\sqrt{1-\xi}\right)\,;\,\, \xi=\frac{2P_0\chi_Q}{n_Q^2}\,.
\end{eqnarray}
where $\tilde{\mu}_e$ is the electron chemical potential, $n_Q(\mu_q,m_s)$ and $\chi_Q(\mu_q,m_s)$ represent the quark charge and quark susceptibility (both slowly varying in the mixed phase) and $P_0$ is the pressure of quarks without electrons, which varies considerably in the mixed phase. These generic relations can be concretely implemented in any specific model of quark matter, so long as we work to second order in the (small) electron chemical potential. Just as for the core, we adopt the Bag model EoS to describe the crust strangelets with ${\cal O}(m_s^4)$ corrections, from which it follows that to the same order, $\mu_e/\mu_q\approx 0.05$, $n_Q(\mu_q)=m_s^2\,\mu_q/(2\pi^2)$ and $\chi_Q(\mu_q)=2\mu_q^2/\pi^2$. Homogeneous quark matter gives way to the mixed phase crust at a radius $r$=$r_c$ where $\xi|_{r=r_c}$=1, i.e, $\tilde{\mu}_e|_{r=r_c}$=$m_s^2(12\mu_q^2-m_s^2)/48\mu_q^3$. We construct the crust as an overlying layer that begins at $\mu_q$=$\mu_{\rm crit}$~\footnote{We include the slight difference between inside and outside pressure for the strangelets due to surface tension in our numerical results. This leads to a slightly higher chemical potential for the strangelet compared to $\mu_{\rm crit}$.}. Hydrostatic equilibrium yields the crustal profile for $r\geq r_c$: 
\begin{equation}
\mu_e(r)=\frac{n_Q}{\chi_Q}\left(1-\sqrt{\frac{2GM}{R^2}\frac{\chi_Q\epsilon_0}{n_Q^2}(r-r_c)}\right)
\end{equation}
where $\epsilon_0$ is the pressure of 3-flavor interacting quark matter without electrons, taken here for the same Bag model parameters as above. For the crust, we take into account the Debye screening effect which preferentially fractionates quark matter into strangelets with size $R_{\ast}\sim\lambda_D$ the Debye screening length. The analysis in~\cite{Alford:2006bx} shows that for surface tension below a critical value, there is an optimal strangelet size $R_{\ast}\approx y\lambda_D$ with $\lambda_D$=$1/\sqrt{4\pi\alpha_e\chi_Q}$ where $\alpha_e$ is the fine structure constant and the dimensionless parameter $y$ is in the range 1.60-2.77. Since the phase fraction $f$ of the strangelets turns out to be quite low, we work in the approximation of isolated strangelets and ignore the possibility of Wigner-Seitz cells and lattice structures for the mixed phase, but in principle these can be studied using the results of~\cite{Alford:2008ge}. The energy density of the mixed phase crust is contributed mainly by the strangelets
\begin{equation}
\epsilon_{\rm crust}=f\epsilon_0\,; \quad f = \frac{\tilde{\mu}_e^3}{3\pi^2n_Q}\left(1-\frac{\chi_Q\tilde{\mu}_e}{n_Q}\right)^{-1}
\label{emix}
\end{equation}
Eqs.(\ref{pmix}) and (\ref{emix}) form a parametric EoS for the crust that is smoothly joined to the underlying homogeneous core at $\xi$=1. The Debye length, and hence the strangelet size is almost constant across the crust, while the strangelet fraction $f$ changes as $\tilde{\mu}_e$ changes with $P_0$. The models with crust have their surface at $\mu_q$=$\mu_{\rm crit}$, so they have very nearly the same mass and radius as the bare strange star models. Plausible values of some parameters describing crust properties for a strange quark star are shown in Table~\ref{Table:crust} below.

\begin{table}[h]
\centering
\begin{ruledtabular}
\begin{tabular}{c c c c c c c c c c c}
$ \mu_{\rm crit}$(MeV) & $m_s$(MeV) & $Z/A$ & $R_{\ast}$(fm)& $\Delta R$(m) & Shear(keV/fm$^3$) &\\ 
\hline
280& 100& 0.039& 8.32& 23.01 & 1.04 \\
290& 150&  0.079& 7.91& 94.44& 87.5\\
300& 200&  0.125& 7.87& 236.66& 1699 \\
\end{tabular}
\end{ruledtabular}
\caption{\protect Numerical values of the charge-to-baryon ratio ($Z/A$), strangelet size $R_{\ast}$, crust thickness $\Delta R$  and Shear modulus at the base of the crust for various combinations of $\mu_{\rm crit}$ and $m_s$ as computed from the interacting Bag model EoS. Mass and radius values for row 1: 2.236 $M_{\odot}$, 12.327 km; row 2: 1.755 $M_{\odot}$, 10.290 km; row 3: 1.140$M_{\odot}$, 7.895 km.}
\label{Table:crust}
\end{table}
\subsection{Neutron Stars}

 {\it Parameterized EoS for the core and crust:} As for strange stars, it is preferable to avail of the same underlying physical model to describe both the core and crust, so that the interface is obtained on physical grounds rather than adhoc matching. With a unified model, changes in the mode spectrum due to a crust can be identified in a systematic manner. We use the smooth analytical parameterization of~\cite{Haensel:2004nu} which is applicable from the core to the outer crust, terminating at density $\rho\simeq 10^{5}$ g/cc. Below $10^5$ g/cc, the neutron star ocean and atmosphere make no discernible change to the high-frequency modes, so we may define the surface of the star at the termination of the outer crust. The ocean and atmosphere, which support thermal gradients, are important for $g$-modes which are not considered here. Based on a fit to either the FPS or SLy EoS at $\rho\geq 5\times 10^{10}$ g/cc, the EoS of Haensel \& Pichon~\cite{HP} for $10^8\leq \rho \,({\mathrm g/cc})\leq 5\times 10^{10}$ and the BPS EoS~\cite{BPS} for $10^5\leq \rho \,({\mathrm g/cc})\leq 10^{8}$ , the unified EoS is parameterized as a log($P$)-log($\rho$) relation. Defining $\xi={\rm log}(\rho/{\rm (g/cc)})$ and $\zeta={\rm log}(P/{\rm (dyn/cm^2)})$, the analytic representation of the unified EoS that is appropriate for non-rotating neutron stars is 
\begin{eqnarray}
\zeta&=&\frac{a_1+a_2\xi+a_3\xi^3}{1+a_4\xi}f_0(a_5(\xi-a_6))\nonumber\\
&+&(a_7+a_8\xi)f_0(a_9(a_{10}-\xi))\nonumber\\ &+&(a_{11}+a_{12}\xi)f_0(a_{13}(a_{14}-\xi))\nonumber\\
&+&(a_{15}+a_{16}\xi)f_0(a_{17}(a_{18}-\xi))
\end{eqnarray}

where $f_0(x)=\frac{1}{({\rm e}^x+1)}$ and the 18 fit parameters $a_1$-$a_{18}$ are given in Table 1 of~\cite{Haensel:2004nu}, for the FPS and SLy EoS. We use the SLy EoS for our calculations. The crust begins in this model at $\xi=14.22$  (about $1.66\times 10^{14}$ g/cc), and at this density, one switches from fluid oscillation variables $y$ to crust variables $z$ (see Appendix). 

{\it Shear modulus for the neutron star crust:} The shear modulus of the  crust is an essential input to the pulsation equations that determine the eigenfrequencies of the non-radial modes. Typically, the $f$-mode frequency is not expected to change significantly on account of the crust, since the mean density of the star is hardly affected. On the other hand, the dispersion for the acoustic $p$-modes is expected to change, as shown by a leading order local analysis~\cite{McDermott:1988}. Although, we will find that this is not necessarily the case for strange stars with a strangelet crust. In addition, there exist shear-driven oscillations that are localized to the crust, as well as interface modes. The numerical values of all these modes depend on the shear modulus. There are numerical calculations of the shear modulus of the neutron star crust in the literature~\cite{Horowitz:2008xr,Baiko:2011cb}. The commonly used analytic expression for the shear modulus of the solid crust is given in the work of Strohmayer~\cite{Strohmayer} 

\begin{equation}
\mu = \frac{0.1194}{1+1.781(\frac{100}{\Gamma})^2}\frac{n_i(Ze)^2}{a^2}
\label{sheareq}
\end{equation}

where $Z$ is the atomic number of the most stable nucleus (ion), $e$ is electron charge, $a$ is the inter-ion separation and $n_i$ is ion-density. For simplicity, we neglect the effect of the variation in $\Gamma$, the Coulomb parameter, in the crust. $\Gamma$ goes from $\sim$ 200 at the ocean-crust boundary to $10^6$ at the crust-core boundary for cold neutron stars (Fig. 2 in~\cite{Strohmayer}), so except for a small region near the ocean-crust boundary, this turns out to be a good numerical approximation to make in Eq.~(\ref{sheareq}). Given that the crust in our chosen EoS model extends from $10^5$ g/cc to $1.66\times 10^{14}$ g/cc, we will use $Z,n_i$ values in 2 separate regimes. For the density regime $5\times 10^5\leq \rho \,({\rm g/cc}) \leq 1.66\times 10^{14}$, we use the step-like $Z$-values tabulated in~\cite{HP}. With these values of $n_i, Z$ across the crust, we can find $a$, the inter-ion separation which is given by $a=(3/(4\pi\,n_i))^{1/3}$.  Since $\mu$ changes in step-like fashion as we march upward in density up to the drip density $\rho_{\rm drip}=4.33\times 10^{11}$ g/cc, we obtain $n_i$ for each density step by using the relation $Zn_i=n_e$ where $n_e=\mu_e^3/(3\pi^2)$ with $\mu_e$ taken from~\cite{HP}. Beyond drip density, i.e, $4.33\times 10^{11}\leq \rho\leq 1.66\times 10^{14}$ g/cc, we switch to the results of Douchin \& Haensel~\cite{DH}, which is consistent with the parameterized EoS we have chosen. This procedure specifies the shear modulus across the span of the crust, from the ocean-crust boundary, through neutron drip, and down to the bottom of the inner crust (i.e, crust-core boundary). 


\section{III. Pulsation Equations}
 
As mentioned in the introduction, the study of non-radial oscillation modes has a long history, starting with the foundational works of Cowling~\cite{Cow}, Pekeris~\cite{Peker} and Kopal~\cite{Kop}. The procedure for computing the spectrum of adiabatic non-radial oscillation modes of zero-temperature stellar objects in General Relativity was laid out in \cite{Thorne:1967,Detweiler:1985zz}. The perturbations can be classified as polar (spheroidal) or axial (toroidal) depending on the parity of the spherical harmonic functional dependence, and are decoupled for non-rotating stars. For the $p$-modes, the restoring force is the pressure, while for the $f$-mode, it is a mixture of pressure and buoyancy.  The importance of these modes is evident from the fact that they can couple strongly to gravitational waves, which carry away the pulsation energy and damp out these modes on timescales of seconds or less~\cite{Andersson:1997rn}. In addition, their excitation in a protoneutron star can lead to the transfer of kinetic energy to the surrounding environment with observable consequences~\cite{Pons:2003}.  There are also core and surface $g$-modes, driven by composition or temperature gradients, which can be studied within more complicated models of the EoS. This last class of non-radial modes, along with discontinuity $g$-modes, are lower in frequency (10 mHz-10 Hz) and consequently outside the range of Earth-based interferometric detectors. However, the $f$-modes are within the accessible range of $\sim$2-3 kHz for spherical detectors~\cite{Costa:2014fwa}, while the $p$-modes are too high in frequency for current detectors.
 
To solve for the spheroidal modes, we use the relativistic Cowling approximation, which neglects the back-reaction of the perturbed fluid on the gravitational potential, but takes General Relativity into account for the structure and fluid perturbations. The Cowling approximation is expected to make at most 10\% difference to the calculation of the eigenfrequencies of the $p$ and $f$ modes, and is widely used in the literature~\cite{Goldreich:1992,Kastaun:2008jr,Flores:2013yqa,Yoshida:2002vd}. The calculation with the complete linearized system of equations in General Relativity without the Cowling approximation, which also yields the damping times for these modes, will be taken up in a following work. 
The system of 4 fluid equations in General Relativity that we solve for the coupled core and crust are detailed in~\cite{Yoshida:2002vd}, as well as the Appendix of this paper for the sake of completeness. For a 2-component star, the crust eigenfunctions (2 for the fluid displacement and 2 for the tractions) are connected to those in the core through the condition of continuity of the radial displacement and tractions, while the horizontal component of the traction vanishes for an ideal fluid~\cite{McDermott:1988}. Additional boundary conditions are imposed to ensure that the eigenfunctions are regular at the center of the star, that the pulsation amplitude is normalized to unity at the stellar surface and that the Lagrangian fluid displacement vanishes there.

\section{IV. Numerical Results}

\begin{figure}[htbp]
\begin{center}
\includegraphics[height=3.2in,width=2.5in,angle=270]{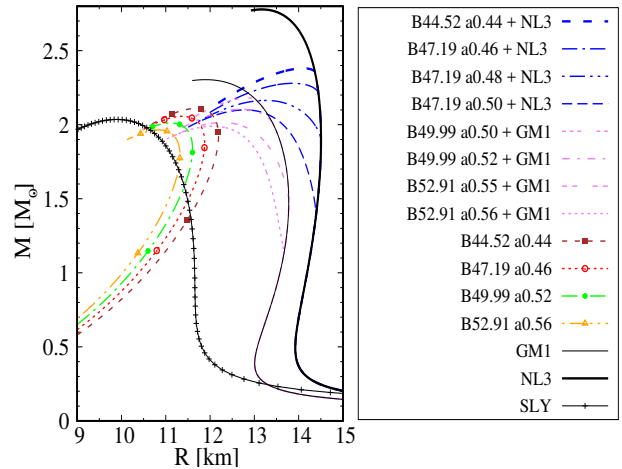}
\caption{Mass-radius curves for bare strange, hadronic and hybrid stars based on the EoS discussed in the text. The values of $B$ and $a_4$ are varied to obtain configurations of bare/hybrid stars of varying maximum mass.}
\label{mvsr}
\end{center}
\end{figure}

Using the compact star models and pulsation formalism mentioned in the previous sections, we compute the frequency of the $f$-mode and lowest $p$-mode. The case of a homogeneous star can be used as a baseline to compare with oscillation modes for a star with a crust. Homogeneous configurations may be broadly categorized into hadronic stars, hybrid stars or bare strange stars. Mass-radius relations for a representative set of hadronic stars, hybrid stars with quark matter, and strange stars are shown in Fig.\,\ref{mvsr}. In addition to the SLy EoS, we have used the GM1~\cite{Glen:91} and NL3~\cite{NL:97} EoS based on relativistic mean field descriptions of dense nuclear matter to construct hadronic stars. For bare strange stars, we employ just the EoS in Eq.\ref{EoS-1} for quark matter with fixed values for $B$ and $a_4$, whereas for hybrid stars we also add the GMI or NL3 parameterizations atop the quark core.  

\begin{figure}[htbp]
\centering
\includegraphics[height=3.2in,width=2.5in,angle=270]{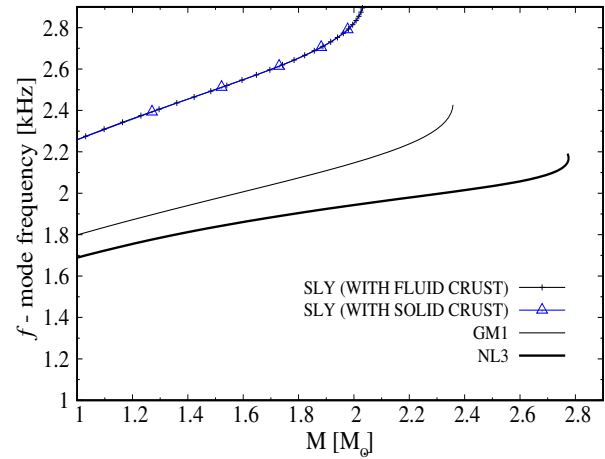}
\caption{The $f$-mode frequencies for a hadronic star for SLy EoS with and without a solid BPS crust, compared with the GM1 and NL3 EoS.}
\label{fslycompare}
\end{figure}

In Fig.~\ref{fslycompare}, we present a comparison of the $f$-mode frequencies for the hadronic star, based on the SLy EoS, with and without a solid (BPS) crust. The legend ``fluid crust" means that we employ Eqs.(20)-(21) of the Appendix, effectively setting the shear modulus to zero at all densities. The legend ``solid crust" means that we employ Eqs.(14)-(17) of the Appendix and include the non-zero shear modulus. It is evident that the addition of a solid crust makes no discernible change to the $f$-mode frequency. The mode frequencies in both cases are in the range of (2.45-2.90) kHz as the mass changes from 1.4$M_{\odot}$ to 2$M_{\odot}$. In comparison to the $f$-mode frequencies for the case of homogeneous stars or hybrid stars studied in~\cite{Flores:2013yqa}, based on the GM1 and NL3 EoS, the SLy EoS gives higher frequencies, as the mean density is higher (smaller radius for the same mass). 

\begin{figure}[htbp]
\centering
\includegraphics[height=3.2in,width=2.5in,angle=270]{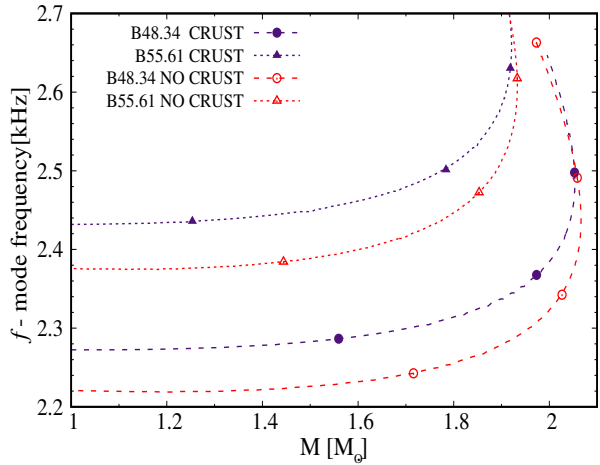}
\caption{The $f$-mode frequencies for a single component (core only) strange quark star compared to the case with a strangelet crust. A large upward shift of 150-200 Hz occurs upon the addition of a strangelet crust.}
\label{fmodecompare}
\end{figure}
\vskip 0.5cm

From Fig.\ref{fmodecompare}, we see that the trend for strange stars is clearly different from hadronic stars. The $f$-mode frequencies are in the (2.20-2.70) kHz range and do not change appreciably with stellar mass up to about 1.5$M_{\odot}$, and begin a steep rise when the maximum mass (typically about 2$M_{\odot}$) is approached. Beyond this value, the $f$-mode frequency shows some back-bending effect, but the star is already subject to radial instabilities at this point. Hadronic stars have $f$-mode frequencies that rise approximately linearly with the square root of the mean density for 1.4$M_{\odot}$-1.8$M_{\odot}$. For homogeneous quark stars in this mass range, the $f$-mode does not scale simply with the square root of the mean density, rather, with a fractional power of the Bag constant. This agrees qualitatively with the results of~\cite{Sotani:2003zc,Flores:2013yqa}. 
Surprisingly, we observe a large upward shift in the $f$-mode frequency of about 200 Hz when a strangelet crust is added to the strange star, an effect that appears in the entire mass range up to the maximum mass. Since the strangelet crust does not change the mean density from the homogeneous case, this large effect is due to the fact that the crust fundamentally changes the surface boundary condition for the $f$-mode from the self-bound case - the pressure, which in the strangelet phase comes essentially from electrons, not quarks, vanishes at zero (electron) density, rather than at high (quark) density. Thus, the crust does have a significant impact on the $f$-mode frequency for strange stars. We find the same effect if we replace the strangelet crust with a thin hadronic crust. This means that the $f$-mode frequency can be useful in determining if there is a large quark core underneath a thin crust. However, it cannot discern between a strangelet and hadronic crust. Essentially, the $f$-mode frequency is sensitive to the nature of matter in the core (self-bound or gravitationally bound) irrespective of the nature of the crust. 

\begin{figure}[!ht]
  \centering
    \includegraphics[height=3.2in,width=2.5in,angle=270]{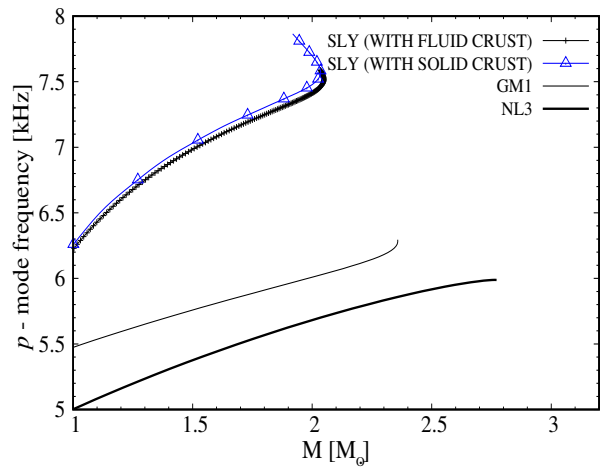}
    \caption{The $p$-mode frequencies for a hadronic star for SLy EoS with and without a solid BPS crust. The acoustic wave speed is affected by the shear modulus of the crust, which implies that $p$-mode frequencies are slightly higher than the homogeneous case.}
    \label{SLY-crustp}
\end{figure}

Turning now to the $p$-mode frequency, Fig.~\ref{SLY-crustp} shows results for hadronic stars with and without a solid crust using the SLy EoS. We see that the $p$-mode frequencies are also higher for the SLy EoS than in the case of the GM1 and NL3 EoS~\cite{Flores:2013yqa}, due to the higher acoustic speed in the former. However, now we also notice that the addition of a solid BPS crust slightly alters the frequencies in the range 1.0$M_{\odot}$-2.0$M_{\odot}$. Since some fraction of the star by volume is in the crust phase, this effect on the acoustic wave speed is a measure of the shear modulus of the crust relative to that of the compression modulus. 

\begin{figure}[!ht]
  \centering
    \includegraphics[height=3.2in,width=2.5in,angle=270]{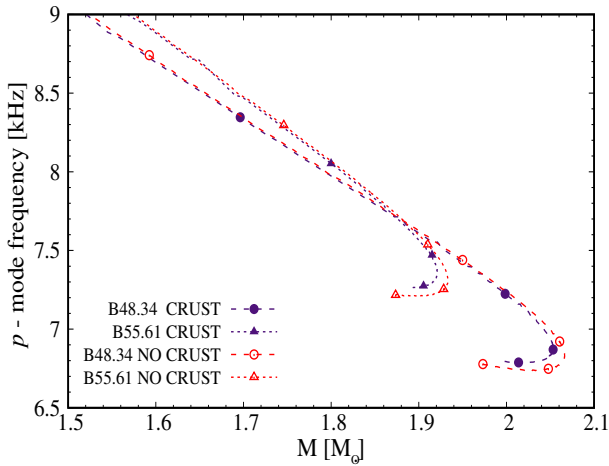}
    \caption{The $p$-mode frequencies for a single component (core only) strange quark star compared to the case with a strangelet crust. The mode frequencies show very little change upon addition of the strangelet crust.}
    \label{pmodecompare}
\end{figure}

Finally, we compare the $p$-modes of strange stars with and without a strangelet crust in Fig.\,\ref{pmodecompare}. Here, we find that the crust makes essentially no difference to the frequencies. At first sight, this may be surprising, since the $p$-modes were certainly affected, albeit slightly, by the shear modulus of the crust in the case of hadronic stars. However, for the strangelet crust, the shear modulus is much smaller than that of a nuclear crust except very close to the interface. The reason is that the shear modulus is proportional to the ionic density (Eq.~(\ref{sheareq})), which is much smaller for the strangelet crust. Since strangelets have approximately the same size but a much higher charge-to-baryon ratio $Z/A$ than neutron-rich nuclei, the strangelets are distributed very sparsely in the crust. This makes the shear modulus quite low, and as such, it does not impact the $p$-mode frequency. Confirming this assertion, if we replace the strangelet crust with a hadronic crust, which has higher shear modulus on average, we again find that the $p$-mode frequency is slightly higher than for bare strange stars. Therefore, the $p$-mode frequency is sensitive to the type of crust, as well as the nature of matter in the core (self-bound or gravitationally bound). 

\section{V. Conclusions}


We studied the non-radial spheroidal modes of oscillation in neutron stars and strange quark stars, including the effects of a BPS crust in the former case and a strangelet or hadronic crust in the latter case. Since non-radial oscillation modes couple to gravitational wave signals, mapping out the mode spectrum as a function of stellar parameters is a useful step in developing gravitational wave templates for pulsating compact stars. The effect of a crust in the two kinds of stars is markedly different. In general, strange quark stars reveal trends in oscillation spectra that do not resemble those of neutron stars. For homogeneous configurations (no crust), the $f$-mode is mostly flat with varying stellar mass in the case of strange stars, but increases linearly in the case of neutron stars. The addition of a BPS crust does not change the $f$-mode frequency for neutron stars, but a strangelet or hadronic crust increases the frequency significantly for quark stars. Since adding a crust to a uniform quark core does not change the average density, we believe this result is unique to self-bound stars with a crust. It is an exciting possibility that gravitational waves could be used to discover large quark matter cores, even if enveloped by a crust of any kind. 

For homogeneous strange stars, the $p$-mode frequency, which depends on the sound speed, decreases with stellar mass at fixed $B$ or even if $B$ and the surface chemical potential $\mu_{\rm crit}$ are varied self-consistently. This is contrary to neutron stars, and due to the self-bound nature of quark matter. Therefore, it appears possible that $p$-modes can also be used to clearly distinguish between neutron and strange stars. However, these frequencies are beyond the reach of current detectors. For a neutron star or bare strange star with a hadronic crust, the $p$-mode frequency increases slightly across the typical mass range, while adding a strangelet crust has no effect on these modes. 

While the strangelet crust faces challenges in explaining the data on QPOs in magnetar oscillations~\cite{Watts:2006mr}, it may support shear ($s$-mode) and interfacial modes ($i$-mode) that are resonantly excited by tidal interaction in binary systems. Tsang et al.~\cite{Tsang:2011ad} have shown that the energy pumped into this mode as a result can shatter the crust of a neutron star, leading to precursor bursts in the short GRB spectrum. It would be interesting to explore if a crust made of strangelets can support such a mode. 

Finally, this work assumes ideal and non-rotating nature of the fluid comprising the compact star. In addition, only the real part of the oscillation frequency is computed. The damping time of these oscillations is necessary in producing templates for gravitational wave observatories. Recent estimates in full general relativity (no Cowling approximation)~\cite{Flores:2017hpb} show that the damping times for $f$-modes are about 1 sec, while that for $p$-modes is 10 seconds or more. That work was restricted to the case of homogeneous non-rotating compact stars, so the effect of the crust and rotation on damping times remains to be calculated along similar lines. Despite the simplicity of the strange star model chosen for this work, our results are significant because they are the first in the field to study the effect of a strangelet crust for strange stars, and to bring out the differences between neutron stars and strange stars with a heterogeneous composition. There remains more work to do in this direction as the era of gravitational waves begins in earnest.

\emph{Acknowledgments.}---P.J. is supported by the U.S. NSF Grant No. PHY 1608959. C. V. F acknowledges the financial support from CAPES. We thank Gautam Rupak, Thomas Kl\"ahn and Jessica Asbell for discussions.

\begin{appendices}
\section{Appendix}

In this appendix, we present the oscillations equations, the boundary and junction conditions that are needed for the numerical evaluation of the spheroidal modes. The pulsation equations were obtained in the framework of the relativistic Cowling approximation.

For the background unperturbed state, we assume the usual Schwarzschild line element:
\begin{eqnarray}
ds^2 = 
     - e^{2 \nu} dt^2 + e^{2 \lambda} dr^2 + r^2 d\theta^2 +
           r^2 \sin^2 \theta d\varphi^2 \, .
\label{metric}
\end{eqnarray}
The total stress-energy tensor is given by
\begin{equation}
T_{\alpha\beta}=\rho \,u_\alpha u_\beta + p \, q_{\alpha\beta} -2 \mu\,
\Sigma_{\alpha\beta} \, ,
\label{EMT}
\end{equation}
where $\mu$ is the isotropic shear modulus, and
a linear relationship between the shear strain and
stress tensor is assumed. $\rho$ and $p$ refer to the mass-energy density and the isotropic pressure, respectively. $u_\alpha$ denotes the fluid 4-velocity and $q_{\alpha\beta}$ is a projection operator with respect to $u$. Finally, $\Sigma_{\alpha\beta}=0$ 
in the strain-free state of the equilibrium configuration. 

After including the fluid perturbations in the equations of energy  and momentum conservation, the system of perturbation
equations can be obtained. In terms of the spheroidal radial perturbations $S_l(r)$ and  $H_l(r)$, the variables in the oscillation equations $z_1$ to $z_4$ are defined as
\begin{eqnarray}
z_1 = S_l (r) \, ,
\end{eqnarray}
\begin{eqnarray}
z_2 = 2 \alpha_1 e^{-\lambda} \frac{d}{dr} \left( r e^{\lambda} S_l (r)
\right) \nonumber
\\ \nonumber
+ \left(\Gamma - \frac{2}{3}\, \alpha_1 \right) \, \left\{
\frac{e^{-\lambda}}{r^2}\,\frac{d}{dr} \left( r^3 e^{\lambda} S_l (r)
\right)\right\} \nonumber
\\- \left\{ l(l+1) H_l (r) \right\} ,
\end{eqnarray}
\begin{eqnarray}
z_3 = H_l (r) \, ,
\end{eqnarray}
\begin{eqnarray}
z_4 = \alpha_1 \, \left( e^{-2 \lambda} r\,\frac{d H_l(r)}{dr} + S_l(r)
\right) \, ,
\end{eqnarray}

where $\alpha_1 = \frac{\mu}{p}$. $z_1-z_4$ obey the following oscillation equations that are solved for in the solid crust:
\begin{eqnarray}
z_1^{'} =  a_{1} z_1 + a_{2}z_2\ 
 +a_{3}z_3
\label{b-eq-s1}
\end{eqnarray}

\begin{eqnarray}
z_2^{'} = a_{4} z_1  
+ a_{5} z_1 + a_{6} z_2
+ a_{7} z_3 
+ a_{8} z_4 
\label{b-eq-s2}
\end{eqnarray}
\begin{eqnarray}
z_3^{'} = a_{9} z_1+ a_{10}
z_4 \, , 
\label{b-eq-s3}
\end{eqnarray}
\begin{eqnarray}
z_4^{'} = a_{11} z_1+ a_{12} z_2 +a_{13} z_3 + a_{14} z_4 \, 
\label{b-eq-s4}
\end{eqnarray}
where the coefficients are 
\begin{eqnarray}
 a_{1}=-\frac{1}{r}\left( 1 +\frac{2 \alpha_2}{\alpha_3} +  r\, \lambdaˆ{'} \right) ,\ 
a_{2}=\frac{1}{r \alpha_3},\ 
a_{3}=\frac{\alpha_2}{r \alpha_3}\, l(l+1) \quad \quad \quad
\nonumber
\\
a_{4}=\frac{1}{r}\left\{ (-3- r\, \lambdaˆ{'}+ \nuˆ{'})^{-1}-e^{2\lambda}\, c_1 \bar\sigma^2 )\,
(1+\rho/p)\,
r\, \nuˆ{'}  \right\} \quad \quad \quad \qquad 
\nonumber
\\
a_{5}=\frac{4\alpha_1}{r \alpha_3}(3\alpha_2+2\alpha_1), 
\
a_{6}=\frac{1}{r}\left( \frac{\rho}{ p} r \nuˆ{'} -4\, \frac{\alpha_1}{\alpha_3} \right) 
 \quad \quad \qquad  \qquad \qquad 
\nonumber
\\
a_{7}=\frac{1}{r}\left\{ (1+\rho/p)\,
r\, \nuˆ{'}-2 \alpha_1 \, \left( 1 + \frac{2 \alpha_2}{\alpha_3} \right)
\right\}\, l(l+1)\,
 \qquad \qquad \qquad
\nonumber
\\
a_{8}= \frac{1}{r}e^{2 \lambda}\, l(l+1)\,
,\
a_{9}= - \frac{1}{r}e^{2 \lambda}\,
,\
a_{10}=\frac{e^{2 \lambda}}{r \alpha_1}\,
 \qquad \qquad \qquad  \quad \quad 
\nonumber
\\
a_{11}=-\frac{1}{r}\left( - (1+\rho/p)\,
r\, \nuˆ{'}+6\Gamma\, \frac{\alpha_1}{\alpha_3}
\right)\, ,
\ a_{12}=-\frac{\alpha_2}{r \alpha_3}\,
 \quad \qquad \qquad
\nonumber
\\
a_{13}=-\frac{1}{r} \left\{
c_1 \bar\sigma^2 (1+\rho/p)\,
r\, \nuˆ{'} +2 \alpha_1-\frac{2 \alpha_1}{\alpha_3}\,
(\alpha_2+\alpha_3)\, l(l+1) \right\}\,
\quad
 \nonumber
 \\
 a_{14}=- \frac{1}{r}(3+ r\, \lambdaˆ{'} - \frac{\rho}{ p}\,
r \nuˆ{'} )
 \qquad \qquad \qquad  \qquad \qquad \qquad \qquad \qquad
 \nonumber
\\ \nonumber
\end{eqnarray}
and the various quantities which appear in the coefficients are
\begin{eqnarray}
\alpha_2 = \Gamma -
\frac{2}{3}\, \alpha_1 \, ,  \ \ \  \alpha_3 = \Gamma +
\frac{4}{3}\, \alpha_1 \, ,
\end{eqnarray}

%

%
%
\begin{eqnarray}
c_1 = \frac{M}{R^3} \, r \, e^{-2 \nu} \, (\nu^{'})^{-1} \,  ,
\end{eqnarray}
where $M$, $R$ are the mass and the radius of the star respectively,
and $\bar\sigma=\sigma\sqrt{R^3/M}$ is the dimensionless frequency.

In the fluid regions, the oscillation equations are
\begin{eqnarray}
y_1^{'} =
b_{15} y_1 + b_{16} y_2 \, ,
\label{b-eq-f1}
\end{eqnarray}
\begin{eqnarray}
y_2^{'} =b_{17} y_1 + b_{18} y_2 \, ,
\label{b-eq-f2}
\end{eqnarray}
where

\begin{eqnarray}
y_1 = S_l (r) \, , \ \ \ \ \
y_2 = (r \nuˆ{'})^{-1} \delta U_l (r) \,=\, c_1\bar\sigma^2 H_l .
\end{eqnarray}
and the coefficients are given by
\begin{eqnarray}
 b_{15}=- \frac{1}{r}\left( 3 - (1+\rho/p)\,
r\, \nuˆ{'}/\Gamma + r\lambda^{} \right) \,,
\qquad \qquad \qquad \qquad
  \nonumber
\\
 \ b_{16} = - \frac{1}{r}\left( (1+\rho/p)\,
r\, \nuˆ{'}/\Gamma - \frac{l(l+1)}{c_1 \bar\sigma^2} \right) ,
  \qquad \qquad \qquad \qquad
  \nonumber
 \\
b_{17}=\frac{1}{r}( e^{2\lambda}\,c_1 \bar\sigma^2 + r A_r ) \, , b_{18}=- \frac{1}{r}(U+r A_r)
 \quad \quad \quad
  \quad \quad 
  \nonumber
\\ \nonumber
\end{eqnarray}
with the Schwarzschild discriminant
\begin{equation}
A_{r}=\frac{1}{\rho + p}\frac{d \rho}{dr} - \frac{1}{\Gamma p}\frac{dp}{dr}
\end{equation}
To calculate the frequency of the modes,
equations (\ref{b-eq-s1})-(\ref{b-eq-s4}) are numerically integrated
in the solid crust, and equations (\ref{b-eq-f1})-
(\ref{b-eq-f2}) in the fluid core.

The outer boundary condition is given
at the stellar surface by $\Delta p = 0$, which reduces to
\begin{eqnarray}
y_1 - y_2 = 0 \, ,
\end{eqnarray}
and the inner boundary condition is the regularity condition at the stellar center
given by
\begin{eqnarray}
c_1 \bar\sigma^2 y_1 - l y_2 = 0 \, .
\end{eqnarray}
Finally the jump conditions  at the interface are given by
\begin{eqnarray}
y_1 = z_1 \, , \ \ \ \ V_1(y_1-y_2) = z_2\, , \ \ \ \ z_4 = 0\, .
\end{eqnarray}

\end{appendices}



\end{document}